\documentstyle[12pt]{article}
\def\al{\alpha}
\def\be{\beta}
\def\ga{\gamma}
\def\de{\delta}
\def\ep{\epsilon}

\def\ze{\zeta}
\def\et{\eta}
\def\th{\theta}

\def\ka{\kappa}
\def\la{\lambda}

\def\rh{\rho}

\def\si{\sigma}

\def\ph{\phi}

\def\ch{\chi}
\def\ps{\psi}
\def\om{\omega}

\def\De{\Delta}

\def\mn{{\mu\nu}}
\def\cl{{\cal L}}
\def\fr#1#2{{{#1} \over {#2}}}
\def\prt{\partial}

\def\expect#1{\langle{#1}\rangle}

\def\half{{\textstyle{1\over 2}}}
\def\frac#1#2{{\textstyle{{#1}\over {#2}}}}

\def\lsim{\mathrel{\rlap{\lower4pt\hbox{\hskip1pt$\sim$}}
    \raise1pt\hbox{$<$}}}
\def\gsim{\mathrel{\rlap{\lower4pt\hbox{\hskip1pt$\sim$}}
    \raise1pt\hbox{$>$}}}
\def\sqr#1#2{{\vcenter{\vbox{\hrule height.#2pt
         \hbox{\vrule width.#2pt height#1pt \kern#1pt
         \vrule width.#2pt}
         \hrule height.#2pt}}}}

\def\Re{\hbox{Re}\,}

\newcommand{\beq}{\begin{equation}}
\newcommand{\eeq}{\end{equation}}
\newcommand{\bea}{\begin{eqnarray}}
\newcommand{\eea}{\end{eqnarray}}
\newcommand{\rf}[1]{(\ref{#1})}

\renewenvironment{thebibliography}[1]
 { \rm
   \begin{list}{\arabic{enumi}.}
    {\usecounter{enumi} \setlength{\parsep}{0pt}
     \setlength{\itemsep}{3pt} \settowidth{\labelwidth}{#1.}
     \sloppy
    }}{\end{list}}

\begin{document}

\begin{center}
{{\large\bf Precision Studies of Relativity in Electrodynamics
\footnote{Presented at
2002 NASA/JPL Workshop for Fundamental Physics in Space,
Dana Point, California, May, 2002.}\\}
\vspace{0.2cm}
Matthew Mewes\\ 
{\small\it Physics Department,
Indiana University,
Bloomington, IN 47405, U.S.A.\\
email: mmewes@indiana.edu\\}}
\vspace{0.3cm}
\parbox{6in}{\small
In this contribution to the
proceedings of the 2002 Workshop
for Fundamental Physics in Space,
a discussion of recent work on
astrophysical and laboratory
tests of Lorentz symmetry in
electrodynamics is presented.
Stringent constraints
are placed on birefringence of
light emitted from
galactic and extragalactic sources.
The prospect of precision
clock-comparison experiments
utilizing resonant cavities
are considered. 
}\end{center}

\vspace{0.4cm}\noindent
{\bf 1. Introduction}\\[.2cm]
In the past, high-precision tests of the
properties of light have played
an important role in the
search for new physics.
Historically, testing the
Lorentz invariance of light
has confirmed special relativity
to a high degree of precision
\cite{mm,kt,classic}.
Many of the traditional experiments
fit into one of two categories.
Michelson-Morley experiments are
designed to test rotational invariance
by searching for anisotropy in the
speed of light.
Kennedy-Thorndike experiments
test boost invariance
by searching for variations
in the speed of light
due to changes in  the velocity
of the laboratory.
In this work, I review
a recent study of extremely precise
tests of Lorentz symmetry in
electrodynamics.
This research was done in collaboration
with Alan Kosteleck\'y.
A detailed discussion can be found
in Ref. \cite{km}.

In recent years, the possibility
that Planck scale physics may
reveal itself at low energies
as small Lorentz violations
has lead to the development
of a general Lorentz-violating
standard-model extension
\cite{ck,cpt01,kle}.
It consists of the
minimal standard model plus small
Lorentz- and CPT-violating terms.
The small violations may originate
from nonzero
vacuum expectation values
of Lorentz tensors in the
underlying theory
\cite{kps}.
Lorentz violations of this type
also arise from noncommutative
field theories
\cite{chklo}.

The extension has provided a
theoretical framework for
a number of high precision
tests of Lorentz symmetry.
To date, experiments involving
hadrons
\cite{kexpt,bexpt,dexpt,ak,ckpvi,bckp},
protons and neutrons
\cite{ccexpt,kla,lh,db,dp},
electrons 
\cite{eexpt,eexpt2},
photons
\cite{km,cfj},
and muons
\cite{muexpt}
have been performed.

A Lorentz-violating extended
electrodynamics can be extracted
from the standard-model extension
\cite{ck}.
In this work, we consider some
experimental consequences of
the extended electrodynamics.
The theory predicts
novel features which
lead to sensitive tests
of Lorentz symmetry.
One unconventional property 
is the birefringence of light.
The observed absence of birefringence
of light emitted from distant sources
leads to tight bounds on some of the
coefficients for Lorentz violation
\cite{km,cfj}.
Some of these bounds are
discussed in Sec. 3.

Another observable consequence of
Lorentz violation is an
orientation and velocity
dependence in the frequencies
of resonant cavities.
This dependence provides the basis
for future clock-comparison
experiments sensitive to the
photon-sector of the standard-model
extension.
Past clock-comparison experiments
have been used to place constraints
on the fermion sector
\cite{ccexpt,kla,lh,db,dp}.
Space-based versions of these
experiments have recently been
considered for precision tests of
Lorentz symmetry on board
the International Space Station
(ISS) and other spacecraft
\cite{spaceexpt}.
Tests for Lorentz violation using
resonant cavities are considered
in Sec. 4.

\vspace{.4cm}\noindent
{\bf 2. Extended Electrodynamics}\\[.2cm]
The photon sector of
the standard-model extension
yields a Lorentz-violating
electrodynamics.
It maintains the usual
gauge invariance
and is covariant under
observer Lorentz transformations.
The Lorentz-violating electrodynamics
includes both CPT-even and -odd terms.
However, the CPT-odd terms are
theoretically undesirable since
they may lead to instabilities
\cite{ck,jk}.
Furthermore, these terms have been
bounded experimentally to
extremely high precision
using polarization measurements
of distant radio galaxies
\cite{cfj}.
Neglecting the CPT-odd terms,
we are left with a CPT-conserving
electrodynamics including small
Lorentz violations.

The CPT-even lagrangian associated
with the Lorentz-violating
electrodynamics is
\cite{ck}
\beq
\cl=-\frac14 F_{\mu\nu}F^{\mu\nu}
-\frac14 (k_F)_{\ka\la\mu\nu}
     F^{\ka\la}F^{\mu\nu}\ ,
\label{lag}
\eeq
where $F_\mn$ is the field strength,
$F_\mn \equiv \prt_\mu A_\nu -\prt_\nu A_\mu$.
The first term is the usual
Maxwell lagrangian.
The second is an unconventional
Lorentz-violating term.
The coefficient for
Lorentz violation,
$(k_F)_{\ka\la\mu\nu}$,
is real and comprised of
19 independent components.
The absence of observed
Lorentz violation requires
$(k_F)_{\ka\la\mu\nu}$
to be small.

It is often convenient to work with the
electric and magnetic fields,
$\vec E$ and $\vec B$, rather
than the vector potential $A^\mu$.
In terms of the usual electric
and magnetic fields,
the lagrangian takes the form
\beq
\cl=\half(\vec E^2-\vec B^2)
+\half \vec E\cdot(\ka_{DE})\cdot\vec E
-\half\vec B\cdot(\ka_{HB})\cdot\vec B
+\vec E\cdot(\ka_{DB})\cdot\vec B\ .
\label{lag2}
\eeq
The real $3\times3$ matrices
$\ka_{DE}$, $\ka_{HB}$
and $\ka_{DB}$
contain the same information as
$(k_F)_{\ka\la\mu\nu}$.
The relationship between the two
notations can be found in Ref.
\cite{km}.
Taking
$\ka_{DE}=\ka_{HB}=\ka_{DB}=0$
in Eq. \rf{lag2} results in the
usual Maxwell lagrangian in terms
of $\vec E$ and $\vec B$.
The parity-even matrices,
$\ka_{DE}$ and $\ka_{HB}$,
are symmetric, while the
parity-odd matrix, $\ka_{DB}$,
has both symmetric and
antisymmetric parts.
The matrices $(\ka_{DE}+\ka_{HB})$
and $\ka_{DB}$ are traceless.
These symmetries leave 11
parity-even and 8 parity-odd
independent components.

The equations of motion for
this lagrangian are 
\beq
\prt_\al{F_\mu}^\al
+(k_F)_{\mu\al\be\ga}\prt^\al F^{\be\ga}
=0\ .
\label{max1}
\eeq
These constitute modified source-free
inhomogeneous Maxwell equations.
The homogeneous Maxwell equations,
\beq
\prt_\mu \widetilde F^{\mn}
\equiv \half\ep^{\mu\nu\ka\la}
\prt_\mu F_{\ka\la}=0\ ,
\label{max1hom}
\eeq
remain unchanged.

An interesting analogy exists between
this theory and the usual situation
in anisotropic media.
Define fields
$\vec D$ and $\vec H$
by the six-dimensional
matrix equation
\beq
\left(
\begin{array}{c} 
\vec D \\ \vec H 
\end{array} 
\right)
=
\left(
\begin{array}{cc} 
 1+\ka_{DE} & \ka_{DB} \\
 \ka_{HE} & 1+\ka_{HB} 
\end{array}
\right)
\left(
\begin{array}{c} 
\vec E \\ \vec B 
\end{array} 
\right)\ ,
\label{DH}
\eeq
with
$\ka_{HE} = -(\ka_{DB})^T$.
Then the modified Maxwell
equations take
the familiar form
\bea
\vec\nabla\times\vec H
- \prt_0 {\vec D} = 0\ ,
& &\quad \vec\nabla\cdot\vec D = 0\ ,
\nonumber \\
\vec\nabla\times\vec E
+ \prt_0 {\vec B} = 0\ ,
& &\quad \vec\nabla\cdot\vec B = 0\ .
\label{max2}
\eea
As a result, the behavior
of electromagnetic fields
in the extended electrodynamics
is very similar to that of
conventional fields in
anisotropic media.

For the purpose of comparing
the sensitivity of various
experiments,
it is convenient to make
the decomposition into four
$3\times 3$ traceless matrices
\bea
(\tilde\ka_{e+})^{jk}
=\half(\ka_{DE}+\ka_{HB})^{jk}\ ,
& &
(\tilde\ka_{e-})^{jk}
=\half(\ka_{DE}-\ka_{HB})^{jk}
   -\frac13\de^{jk}(\ka_{DE})^{ll}\ ,
\nonumber\\
(\tilde\ka_{o+})^{jk}
=\half(\ka_{DB}+\ka_{HE})^{jk}\ ,
& &
(\tilde\ka_{o-})^{jk}
=\half(\ka_{DB}-\ka_{HE})^{jk}\ ,
\label{kappatilde}
\eea
and a single rotationally symmetric
trace component
\beq
\tilde\ka_{\rm tr}
=\frac13(\ka_{DE})^{ll}\ .
\label{kappatrace}
\eeq
The matrices
$\tilde\ka_{e+}$ and
$\tilde\ka_{e-}$ 
and the single coefficient
$\tilde\ka_{\rm tr}$
contain the parity-even
coefficients, while
the matrices
$\tilde\ka_{o+}$ and
$\tilde\ka_{o-}$ 
contain the parity-odd.
The matrix $\tilde\ka_{o+}$
is antisymmetric while the
other three are symmetric.

In terms of this decomposition,
the lagrangian is
\bea
\cl&=&\half[(1+\tilde\ka_{\rm tr})\vec E^2
-(1-\tilde\ka_{\rm tr})\vec B^2]
+\half \vec E\cdot(\tilde\ka_{e+}
+\tilde\ka_{e-})\cdot\vec E
\nonumber\\
&&
-\half\vec B\cdot(\tilde\ka_{e+}
-\tilde\ka_{e-})\cdot\vec B 
+\vec E\cdot(\tilde\ka_{o+}
+\tilde\ka_{o-})\cdot\vec B\ .
\label{lag3}
\eea 
From the form of Eq. \rf{lag3},
it is evident that the component
$\tilde\ka_{\rm tr}$
corresponds to shift in the
effective permittivity $\ep$ and
effective permeability $\mu$ by
$(\ep-1)=-(\mu^{-1}-1)
=\tilde\ka_{\rm tr}$.
Therefore, the effect of a
nonzero $\tilde\ka_{\rm tr}$
is an overall shift in the
speed of light.
This result generalizes to the
nine independent coefficients in
$\tilde\ka_{\rm tr}$,
$\tilde\ka_{e-}$ and
$\tilde\ka_{o+}$.
To leading order,
these can be viewed as a distortion
of the spacetime metric of the form
$\et^\mn\rightarrow\et^\mn+k^\mn$,
where $k^\mn$ is small, real and
symmetric.

Small distortions of this type
are normally unphysical, since
they can be eliminated through
coordinate transformations and
field redefinitions.
However, in the context of the
full standard-model extension,
eliminating these terms from
the photon sector will alter
other sectors and the effects
of such terms can not be removed.
Thus, in experiments where
the properties of light are
compared to the properties of
other matter, these terms
are relevant.
While in experiments sensitive
to the properties of light only,
these nine coefficients are not
expected to appear.
The resonant-cavity based
experiments discussed in
Sec. 4 fall into the first category,
while the astrophysical tests
of Sec. 3 belong to the second.
The tests discussed discussed in
Sec. 3 rely on measurements of
birefringence, which in essence
compares the properties of light
with different polarizations.
Therefore, these tests are only
sensitive to the ten independent
components of $\tilde\ka_{e+}$
and $\tilde\ka_{o-}$.

When reporting bounds on
birefringence it is convenient
to express them in terms of a
ten-dimensional vector $k^a$
containing the ten independent
components of
$\tilde\ka_{e+}$ and $\tilde\ka_{o-}$
\cite{km}.
The relationship between
$\tilde\ka_{e+}$, $\tilde\ka_{o-}$
and $k^a$ is given by
\bea
(\tilde\ka_{e+})^{jk} &=&
-\left(
\begin{array}{ccc}
-(k^3+k^4) & k^5 & k^6 \\
k^5 & k^3 & k^7 \\
k^6 & k^7 & k^4 
\end{array}
\right)\ , 
\nonumber \\
(\tilde\ka_{o-})^{jk} &=&
\left(
\begin{array}{ccc}
2k^2 & -k^9 & k^8 \\
-k^9 & -2k^1 & k^{10} \\
k^8 & k^{10} & 2(k^1-k^2) 
\end{array}
\right)\ .
\eea
Bounds can then be expressed in
terms of $|k^a|\equiv\sqrt{k^ak^a}$,
the magnitude of the vector $k^a$.

\vspace{.4cm}\noindent
{\bf 3. Astrophysical Tests}\\[.2cm]
In order to understand the
effects of Lorentz violation
on the propagation of light,
we begin by considering
plane-wave solutions.
Adopting the ansatz
$F_\mn(x)=F_\mn e^{-ip_\al x^\al}$,
the modified Maxwell equations 
yield an Amp\`ere law
given by the linear equation
\beq
(-\de^{jk} p^2 - p^j p^k 
-2(k_F)^{j \be \ga k} p_\be p_\ga)E^k = 0\ .
\label{ampere}
\eeq
Solving this equation determines
the dispersion relation
\beq
p^0_\pm=(1+\rh\pm\si)|\vec p|\ ,
\label{dispersion}
\eeq
and electric field
\beq
\vec E_\pm\propto(\sin\xi,\pm1-\cos\xi,0)
+O(k_F)\ .
\label{E_field}
\eeq
To leading order,
the quantities $\rh$,
$\si\sin\xi$ and $\si\cos\xi$
are linear combinations of
$(k_F)_{\ka\la\mu\nu}$
and depend on $\hat v$,
the direction of propagation.
One unconventional feature
of these solutions is the birefringence
of light in the absence of matter.
In the conventional case,
this behavior is commonly found
in the presence of anisotropic media.

The general vacuum solution is
a linear combination of the
$\vec E_+$ and $\vec E_-$.
For nonzero $\si$, these solutions
obey different dispersion relations.
As a result, they propagate
at slightly different velocities.
At leading order, the difference
in the velocities is given by
\beq
\De v \equiv v_+-v_- = 2\si\ .
\label{dev}
\eeq
For light propagating over
astrophysical distances, this
tiny difference may become apparent.

As can be seen from the
above solutions,
birefringence depends on
the linear combination
$\si\sin\xi$ and $\si\cos\xi$.
As expected, these only contain the ten
independent coefficients which appear
in $\tilde\ka_{e+}$ and $\tilde\ka_{o-}$.
Expressions for 
$\si\sin\xi$ and $\si\cos\xi$
in terms of these ten
independent coefficients and
the direction of propagation
can be found in the literature
\cite{km}.

Next, we consider two observable effects
stemming from the birefringence.
The first is the difference
in arrival time of two modes in
unpolarized light.
The second is the change
in polarization of polarized light
emitted from distant sources.

\vspace{.4cm}\noindent
{\it 3.1. Pulse Dispersion}\\[.2cm]
For a source producing
relatively unpolarized light,
the components $\vec E_\pm$ associated
with each mode will be comparable.
For light produced at a given
instant, the difference in velocity 
will induce a difference in the
observed arrival time of the two modes
given by $\De t \simeq \De v L$,
where $L$ is the distance to the source.

To make use of this idea,
we consider sources that produce
radiation with rapidly changing time
structure.
Sources producing narrow pulses
of radiation such as pulsars and
gamma-ray bursts are ideal.
For sources of this type,
the pulse can be regarded as the
superposition of two independent
pulses, one for each mode.
As the pulse propagates, the
difference in velocity will cause
the two pulses to separate.
A signal for Lorentz violation would
manifest itself as the observation
of two sequential pulses of similar
structure.
The pulses would be linearly polarized
at mutually orthogonal polarization angles.
This type of double pulse
has not yet been observed.

\begin{table}[b!]
\begin{center}
\begin{tabular}{|l||rl|rl|}
\hline
\multicolumn{1}{|c||}{Source} &
\multicolumn{2}{c|}{$L$} &
\multicolumn{2}{c|}{$w_o$} \\ 
\hline \hline
GRB\,971214 \cite{kulk2,batse}   & 2.2 &Gpc& 50  &s    \\
GRB\,990123 \cite{batse,kulk1}     & 1.9 &Gpc& 100 &s    \\
GRB\,980329 \cite{batse,frutcher}     & 2.3 &Gpc& 50  &s    \\
GRB\,990510 \cite{batse,gcn324}     & 1.9 &Gpc& 100 &s    \\
GRB\,000301C \cite{gcn568,gcn605}    & 2.0 &Gpc& 10  &s    \\
PSR\,J1959+2048 \cite{psrcat} & 1.5 &kpc& 64  &$\mu$s    \\
PSR\,J1939+2134 \cite{psrcat} & 3.6 &kpc& 190 &$\mu$s    \\
PSR\,J1824-2452 \cite{psrcat} & 5.5 &kpc& 300 &$\mu$s    \\
PSR\,J2129+1210E \cite{psrcat}& 10.0 &kpc& 1.4 &ms    \\
PSR\,J1748-2446A \cite{psrcat}& 7.1 &kpc& 1.3 &ms    \\
PSR\,J1312+1810 \cite{psrcat} & 19.0 &kpc& 4.4 &ms    \\
PSR\,J0613-0200 \cite{psrcat} & 2.2 &kpc& 1.4 &ms    \\
PSR\,J1045-4509 \cite{psrcat} & 3.2 &kpc& 2.2 &ms    \\
PSR\,J0534+2200 \cite{psrcat,gpp2} & 2.0 &kpc& 10  &$\mu$s    \\
PSR\,J1939+2134 \cite{psrcat,gpp1} & 3.6 &kpc& 5   &$\mu$s    \\
\hline
\end{tabular}
\end{center}
\caption{Source data for velocity constraints.}
\end{table}

Single pulse measurements can be used
to place bounds on Lorentz violation.
Suppose a source produces a pulse
with a characteristic width $w_s$.
As the pulse propagates, the two
modes spread apart and the width of the
pulse will increase.
The observed width can be estimated
as $w_o \simeq w_s+\De t$.
Therefore, observations of $w_o$
place conservative bounds on
$\De t \simeq \De v L \simeq 2\si L$.
The resulting bound on $\si$ constrains
the ten-dimensional parameter space of
$\tilde\ka_{e+}$ and $\tilde\ka_{o-}$.
Since a single source constrains only
one degree of freedom, a minimum of ten
sources located at different positions
on the sky are required to fully constrain
the ten coefficients.

Table 1 lists a sample of 16 sources
consisting of gamma-ray bursts and pulsars,
as well as their distances and pulse widths.
Each width places a bound on $\si$ for
that particular source.
Combining these bounds using a
method described in Ref.
\cite{km} constrains
the ten-dimensional parameter space.
At the 90\% confidence level, we
obtain a bound of
$|k^a| < 3 \times 10^{-16}$ on
the coefficients for Lorentz violation.

\vspace{.4cm}\noindent
{\it 3.2. Spectropolarimetry}\\[.2cm]
In this subsection, we consider the
effect of Lorentz violation
on polarized light.
Decomposing a general electric field
into its birefringent components,
we write
\beq
\vec E(x) = (\vec E_+ e^{-ip^0_+t}+
\vec E_- e^{-ip^0_-t})e^{i\vec p \cdot\vec x}\ .
\eeq
Each of the components propagates
with different phase velocity.
A change in the relative phase
results from this difference.
The shift in relative phase
is given by
\beq
\De\ph
= (p^0_+-p^0_-)t 
\simeq 4\pi\si L/\la\ ,
\eeq
where $L$ is the distance to the
source and $\la$ is the wavelength
of the light.
The change in relative phase
results in a change in the
polarization as the radiation
propagates.

The $L/\la$ dependence suggests
the effect is larger for more
distant sources and shorter
wavelengths.
Recent spectropolarimetry of distant
galaxies at wavelengths ranging from
infrared to ultraviolet
has made it possible to achieve values
of $L/\la$ greater than $10^{31}$.
Measured polarization parameters
are typically order 1.
Therefore, we expect an experimental
sensitivity of $10^{-31}$ or better
to components of $(k_F)_{\ka\la\mu\nu}$.

In general, plane waves
are elliptically polarized.
The polarization ellipse can
be parameterized with angles 
$\psi$, which characterizes the
orientation of the ellipse, and
$\chi=\pm\arctan
\frac{\rm minor\ axis}{\rm major\ axis}$,
which describes the shape of the
ellipse and helicity of the wave.
The phase change, $\De\ph$, results
in a change in both $\psi$ and $\chi$.
However, measurements of $\chi$ are not
commonly found in the literature.
Focusing our attention on $\psi$,
we seek an expression for
$\de\psi=\psi-\psi_0$,
the difference between $\psi$ at
two wavelengths, $\la$ and $\la_0$.
We find
\cite{km}
\beq
\de\psi
=\half\tan^{-1}{\fr
{\sin\tilde\xi\cos\ze_0
+\cos\tilde\xi\sin\ze_0\cos(\de\ph-\ph_0)}
{\cos\tilde\xi\cos\ze_0
-\sin\tilde\xi\sin\ze_0\cos(\de\ph-\ph_0)}},
\label{dpsi}
\eeq
where we have defined
$\de\ph=4\pi\si(L/\la-L/\la_0)$,
$\tilde\xi=\xi-2\psi_0$ and 
$\ph_0\equiv \tan^{-1}
(\tan2\ch_0/\sin\tilde\xi)$,
$\ze_0\equiv \cos^{-1}
(\cos2\ch_0\cos\tilde\xi)$.
The polarization at $\la_0$ is given
by the polarization angles
$\ps_0$ and $\ch_0$.

The idea is to fit
existing spectropolarimetric
data to Eq. \rf{dpsi}.
Under the reasonable assumption
that the polarization of the
light when emitted is relatively
constant over the relevant wavelengths,
any measured wavelength dependence
in the polarization is due to Lorentz
violation.

\begin{table}[t]
\begin{center}
\begin{tabular}{|l||c|c|c|}
\hline
\multicolumn{1}{|c||}{Source} 
& $L_{eff}$~(Gpc)
& $10^{30}L_{eff}/\la$
& $\log_{10}\si$ \\ 
\hline \hline
IC 5063 \cite{hough}                      & 0.04 & 0.56 - 2.8 & -30.8 \\
3A 0557-383 \cite{brindle}                & 0.12 & 2.2 - 8.5  & -31.2 \\
IRAS 18325-5925 \cite{brindle}            & 0.07 & 1.0 - 4.9  & -31.0 \\
IRAS 19580-1818 \cite{brindle}            & 0.14 & 1.8 - 9.3  & -31.0 \\
3C 324 \cite{cimatti465}                  & 2.44 & 82 - 180   & -32.3 \\
3C 256 \cite{dey}                         & 3.04 & 110 - 220  & -32.4 \\
3C 356 \cite{cimatti476}                  & 2.30 & 78 - 170   & -32.3 \\
F J084044.5+363328 \cite{brothertonfirst} & 2.49 & 88 - 170   & -32.4 \\
F J155633.8+351758 \cite{brothertonfirst} & 2.75 & 99 - 160   & -32.4 \\
3CR 68.1 \cite{brotherton}                & 2.48 & 84 - 180   & -32.4 \\
QSO J2359-1241 \cite{brothertonqso}       & 2.01 & 110 - 120  & -31.2 \\
3C 234 \cite{kishimoto}                   & 0.61 & 55 - 81    & -31.7 \\
4C 40.36 \cite{vernet}                    & 3.35 & 120 - 260  & -32.4 \\
4C 48.48 \cite{vernet}                    & 3.40 & 120 - 260  & -32.4 \\
IAU 0211-122 \cite{vernet}                & 3.40 & 120 - 260  & -32.4 \\
IAU 0828+193 \cite{vernet}                & 3.53 & 130 - 270  & -32.4 \\
\hline
\end{tabular}
\end{center}
\caption{Source data for polarization constraints.}
\end{table}

Table 2 lists 16 sources with
published polarimetric data
with wavelengths ranging from
400 to 2200 nm.
In this table, the effective
distance $L_{eff}$ is listed
which takes cosmological redshift
of the light into account.
Using a fitting procedure
described in Ref. \cite{km},
we obtain a bound on $\si$ for
each source.
Combining these bounds in the
same manner as in the
pulse-dispersion case,
a constraint on the ten-dimensional
parameter space is found.
At the 90\% confidence level,
we obtain a bound of
$|k^a| < 2 \times 10^{-32}$ on 
the coefficients for
Lorentz violation responsible
for birefringence.

\vspace{.4cm}\noindent
{\bf 4. Resonant Cavities}\\[.2cm]
Clock-comparison experiments have
proved to be some of the most
sensitive tests of Lorentz symmetry
\cite{ccexpt,kla,lh,db,dp}.
The frequencies of these clocks 
are typically atomic Zeeman transitions.
Lorentz violation causes these
frequencies to vary with changes
in orientation or velocity of the clock.
Experiments searching for a variation
due to the rotational motion of the
Earth have placed stringent
bounds on Lorentz violation
in the fermion sectors of the
standard-model extension.

Modern versions of the Michelson-Morley
and Kennedy-Thorndike experiments
utilize resonating
electromagnetic cavities
\cite{bh,hh,brax}.
Resonant cavities serve as clocks
in clock-comparison experiments
which are sensitive to Lorentz
violation in the photon sector.
These experiments search for 
a variation in the resonant frequency
of a cavity as its orientation or
velocity changes.
For a typical Earth-based experiment,
the variation in resonant frequency
occurs at harmonics of the Earth's
sidereal frequency,
$\om_\oplus \simeq 2\pi /$(23 hr 56 min).
Due to the orbital motion of the Earth,
the variation may also contain
annual components.

The ISS and other spacecraft
provide interesting platforms for future
clock-comparison experiments.
The orbital properties of the spacecraft
may result in radically different behavior.
For example, the orbital period of
the ISS is about 92 min.
The comparable period for an Earth-based
experiment is the Earth's sidereal period.
This suggests a significant reduction is
data-acquisition time for a space-based
experiment compared to its Earth-based
counterpart.

We begin our discussion by considering
the effects of Lorentz violation
on the resonant frequency of cavities.
We then consider two classes of
cavities, optical and microwave,
which are currently under development
for precision tests of relativity.

\vspace{.4cm}\noindent
{\it 4.1 General Considerations}\\[.2cm]
The quantity of interest is the
fractional resonant-frequency
shift $\de\nu/\nu$.
Consider a harmonic system satisfying
the Maxwell equations \rf{max2}.
Suppose $\vec E_0$, $\vec B_0$,
$\vec D_0$ and $\vec H_0$ are the
conventional solutions with resonant
angular frequency $\om_0$.
Let $\vec E$, $\vec B$, $\vec D$ and
$\vec H$ be solutions for nonzero
$(k_F)_{\ka\la\mu\nu}$ with angular
frequency $\om$.
Manipulating the Maxwell equations
for both sets of fields,
we obtain the expression
\bea
\fr{\de\nu}\nu=
\fr{\om-\om_0}{\om_0}&=&
-\left(\int_V d^3x
\bigl(\vec E_0^*\cdot\vec D
+\vec H_0^*\cdot\vec B\bigr)\right)^{-1}
\nonumber\\
&&
\times\int_V d^3x
\left(\vec E_0^*\cdot\vec D
-\vec D_0^*\cdot\vec E
-\vec B_0^*\cdot\vec H
+\vec H_0^*\cdot\vec B
\right. 
\nonumber\\
&&
\left.
\qquad
\qquad
-i\om_0^{-1}\vec\nabla\cdot
(\vec H_0^*\times\vec E
-\vec E_0^*\times\vec H)
\right) ,
\label{dnu}
\eea
where the integrals are over the
volume $V$ of the cavity.

Note that the divergence term in 
Eq. \rf{dnu} results in a surface
integral over the boundary of $V$.
In many situations, we can neglect
such boundary terms.
For example, neglecting Lorentz
violations in other sectors,
the fields vanish inside a perfect
conductor, by usual arguments.
Idealizing the walls of the cavity
as a perfect conductor,
the Faraday equation
$\vec\nabla\times\vec E
+ \prt_0 {\vec B} = 0$,
implies the tangential component of
$\vec E$ vanishes at the surface.
In this scenario, the divergence
term in Eq. \rf{dnu} is zero.

Using Eq. \rf{dnu}, we can
find the frequency shift
perturbatively in terms of
the conventional solutions.
For a cavity void of matter,
we have
$\vec E_0 =\vec D_0$ and
$\vec B_0 =\vec H_0$.
The constitutive relations \rf{DH}
give the approximate equalities
\beq
\vec D-\vec E \simeq\ka_{DE}
\cdot\vec E_0+\ka_{DB}\cdot\vec B_0\ ,
\quad
\vec H-\vec B \simeq\ka_{HE}
\cdot\vec E_0+\ka_{HB}\cdot\vec B_0\ .
\eeq
With these relations and
the vanishing of the boundary
term, the leading order
fractional frequency shift is
\beq
\fr{\de\nu}\nu 
= -\fr1{4\expect U}\int_V d^3x \,
\left(\vec E_0^* \cdot \ka_{DE} \cdot \vec E_0
-\vec B_0^* \cdot \ka_{HB} \cdot \vec B_0
+2\Re\bigl(\vec E_0^*\cdot\ka_{DB}
\cdot\vec B_0\bigr)\,\right)\ ,
\label{dnu1}
\eeq
where 
$\expect U=\int_V d^3x\,
(|\vec E_0|^2+|\vec B_0|^2)/4$
is the time-averaged energy
stored in the unperturbed cavity.
Note that $\de\nu/\nu$ is real,
indicating that the $Q$ factor
of the cavity remains unaffected
by Lorentz violation at leading order.

The integrals in Eq. \rf{dnu1}
are most readily carried out
in a frame at rest with respect
to the laboratory.
Since the laboratory frame
is not inertial in general,
the laboratory-frame coefficients
$(\ka_{DE})_{\rm lab}^{jk}$,
$(\ka_{DB})_{\rm lab}^{jk}$ and
$(\ka_{HB})_{\rm lab}^{jk}$
are not constant.
However, using observer covariance,
the laboratory-frame coefficients
can be related to the
coefficients in an inertial
frame through observer Lorentz
transformations.

There are many logical
candidates for an inertial frame.
For our purposes,
a Sun-centered celestial
equatorial frame will suffice.
The coefficients in this frame
$(\ka_{DE})^{JK}$,
$(\ka_{HB})^{JK}$ and
$(\ka_{DB})^{JK}$ can be
regarded as constant.
The relative smallness of
the velocity of the Earth
in this frame,
$\be_\oplus \approx 10^{-4}$,
implies it is usually sufficient
to expand the transformation
in powers of the velocity $\be$.
To order $\be$,
the relation between the
laboratory-frame coefficients
and the Sun-frame coefficients
is given by
\bea
(\ka_{DE})_{\rm lab}^{jk}
&=&T_0^{jkJK}(\ka_{DE})^{JK}
-T_1^{(jk)JK}(\ka_{DB})^{JK}\ ,
\nonumber\\
(\ka_{HB})_{\rm lab}^{jk}
&=&T_0^{jkJK}(\ka_{HB})^{JK}
-T_1^{(jk)KJ}(\ka_{DB})^{JK}\ ,
\nonumber\\
(\ka_{DB})_{\rm lab}^{jk}
&=&T_0^{jkJK}(\ka_{DB})^{JK}
+T_1^{kjJK}(\ka_{DE})^{JK}
+T_1^{jkJK}(\ka_{HB})^{JK}\ ,
\label{trans}
\eea
with
$T_0^{jkJK}\equiv R^{jJ}R^{kK}$
and
$T_1^{jkJK}\equiv R^{jP}R^{kJ}\ep^{KPQ} \be^Q$,
where
$R^{jJ}$ is the rotation
from the Sun frame to the
laboratory frame, and
$\be^Q$ is the velocity
of the laboratory in the
Sun frame.
The tensor $T_0$ is a rotation,
while $T_1$ is a leading-order
boost contribution.
Although the terms involving
$T_1$ are suppressed by $\be$,
they access distinct combinations
of coefficients and can introduce
different time dependence, 
which may lead to fundamentally
different tests.

\vspace{.4cm}\noindent
{\it 4.2 Optical Cavities}\\[.2cm]
Recent examples of modern
Michelson-Morley and
Kennedy-Thorndike experiments
based on optical cavities
include Refs.
\cite{bh,hh,brax}.
The basic setup of these
experiments consists of
a pair of stabilized lasers.
One laser is stabilized by
an optical cavity.
The second laser is stabilized by
a molecular transition which
in the classical analysis is
assumed to be insensitive
to Lorentz violations.
This laser serves as a reference
frequency.
The beat frequency of the combined
signal is analyzed for a variation
due to a change in the orientation
or velocity of the cavity.

The sensitivities achieved in these
experiments are typically on the
order of $10^{-13}$ to $\de\nu/\nu$.
Analyzing these experiments
in the context of the extended
electrodynamics should therefore
yield bounds on components of
$(k_F)_{\ka\la\mu\nu}$ at the
level of $10^{-13}$.

Regarding these cavities as
two parallel planar reflecting
surface, the usual solutions
can be approximated as standing waves.
In a reference frame at
rest in the laboratory,
we take
\bea
\vec E_0(x)&=&\vec E_0
\cos(\om_0\hat N\cdot\vec x+\ph)e^{-i\om_0t}\ ,
\nonumber\\
\vec B_0(x)&=&i\hat N\times\vec E_0
\sin(\om_0\hat N\cdot\vec x+\ph)e^{-i\om_0t}\ ,
\label{reswave}
\eea
where $\hat N$ is a unit
vector pointing along the
length of the cavity, 
$\ph$ is a phase,
and $\vec E_0$ is a
vector perpendicular to $\hat N$ 
that specifies the polarization.
The resonant frequencies of the
conventional solutions are
given by $\om_0=\pi m / l$,
where $m$ is an integer
and $l$ is the separation of the
reflecting surfaces.

Substituting this solution into
Eq. \rf{dnu1} yields
the fractional frequency shift:
\beq
\fr{\de\nu}{\nu}
=-\fr1 {2|\vec E_0|^2}
\bigl[\vec E_0^*\cdot
(\ka_{DE})_{\rm lab}\cdot\vec E_0
-(\hat N \times \vec E_0^*)
\cdot(\ka_{HB})_{\rm lab}\cdot
(\hat N \times \vec E_0)\bigr]\ .
\label{dnuopt}
\eeq
This result depends on the orientation
of the cavity in the laboratory
and the polarization of the light.
Transforming the laboratory-frame
coefficients to the Sun-frame,
using Eq. \rf{trans},
introduces variations in the
frequency shift due to the
motion of the lab.

Consider a Earth-based laboratory.
The transformation \rf{trans} 
includes variations at the Earth's
sidereal and orbital frequencies.
The orbital frequency components are
a result of a boost, and are therefore
suppressed relative to the purely
rotational contributions.
Consequently, the resonant frequency
fluctuates at $\om_\oplus$
and the second harmonic $2\om_\oplus$,
along with suppressed oscillations
associated with the annual variation
in the Earth's velocity.

Different experimental
configurations result in
different sensitivities to
the coefficients
$(k_F)_{\ka\la\mu\nu}$,
and can result in different
frequencies in frequencies in the
variations of $\de\nu/\nu$.

As an example, consider
a cavity positioned horizontally
in the laboratory with vertical
polarization.
Let $\th$ be an angle specifying
the orientation of the cavity
in the horizontal plane.
The frequency shift takes the form
\beq
\fr{\de\nu}\nu = A + B\sin2{\th} +C\cos2{\th} , 
\label{ffs}
\eeq
where
\bea
A &=& A_0+A_1\sin\om_\oplus T_\oplus
 +A_2\cos\om_\oplus T_\oplus 
 +A_3\sin2\om_\oplus T_\oplus
 +A_4\cos2\om_\oplus T_\oplus\ , 
\nonumber \\
B &=& B_0+B_1\sin\om_\oplus T_\oplus
 +B_2\cos\om_\oplus T_\oplus 
 +B_3\sin2\om_\oplus T_\oplus
 +B_4\cos2\om_\oplus T_\oplus\ , 
\nonumber \\
C &=& C_0+C_1\sin\om_\oplus T_\oplus
 +C_2\cos\om_\oplus T_\oplus 
 +C_3\sin2\om_\oplus T_\oplus
 +C_4\cos2\om_\oplus T_\oplus\ .
\label{dnu11}
\eea
The quantities 
$A_{0,1,2,3,4}$, $B_{0,1,2,3,4}$,
and $C_{0,1,2,3,4}$ 
are linear in the coefficients
for Lorentz violation and depend
on the latitude of the laboratory.

From Eq. \rf{ffs}, we see that
one possible strategy for searches
for Lorentz-violation would
be to rapidly rotate the cavity
in the laboratory and search for
variations at the harmonics of the
cavity rotation frequency.
This is the method used in the
experiment of Brillet and Hall
\cite{bh}.
It has been estimated that their
analysis constrains one combination
of coefficients to about a part
in $10^{15}$ \cite{km}.

Hills and Hall performed an
experiment with the cavity
fixed in the laboratory
\cite{hh}.
A bound is placed on the
sidereal variation on the
order of $10^{-13}$.
We see from Eqs. \rf{ffs}
and \rf{dnu11} that
this constrains some combination
of the coefficients
$A_1$, $A_2$, $B_1$, $B_2$,
$C_1$, and  $C_2$.

A similar experiment has recently
been performed by Braxmaier {\it et al.}
\cite{brax}.
Their analysis focuses on
variations due to the orbital
motion of the Earth.
In the present context,
this corresponds to $\be$
suppressed terms arising from
the leading order boost
contributions in the
transformation \rf{trans}.
They achieve fractional-frequency
sensitivity of $4.8\pm5.3\times10^{-12}$,
which leads to an estimated
constraint on a combination
of coefficients on the order of $10^{-8}$.

It should be noted that $\be$ suppressed
terms involve parity-odd coefficients,
while the unsuppressed terms are only
sensitive to parity-even coefficients.
Therefore, consideration of these
terms seems worthwhile even at
reduced sensitivity.

The above experiments place loose
constraint on three combinations
of coefficients.
It is likely that reanalyzing
these experiments in terms
of the standard-model extension
would place constraints on more
combinations at similar levels.

\vspace{.4cm}\noindent
{\it 4.3 Microwave Cavities}\\[.2cm]
Microwave-cavity oscillators
are among the most stable clocks.
Cavities constructed of superconducting
niobium have achieved frequency
stabilities of $3\times10^{-16}$.
In an effort to perform improved tests
of relativity, superconducting microwave
oscillators are being developed
by the SUMO project for use on
upcoming ISS missions
\cite{sumo}.

Equation \rf{dnu1} can be applied
to cavities of any geometry and
operated in any mode.
Here we consider a cylindrical
cavity of circular cross section,
operated in the fundamental
TM$_{010}$ mode.
The integrals in Eq. \rf{dnu1}
are easily carried out in a
frame fixed to the cavity with
its 3 axis along the symmetry axis.
In terms of coefficients in
the cavity-fixed frame,
the fractional frequency shift is
\beq
\fr{\de\nu}\nu =
-\half (\ka_{DE})_{\rm cav}^{33} 
+\frac14[(\ka_{HB})_{\rm cav}^{11}
+(\ka_{HB})_{\rm cav}^{22}]\ .
\eeq
It is not difficult to
generalize this expression to
an arbitrary laboratory frame
in which the cavities symmetry
axis points in a direction
specified by a unit vector $\hat N$.
The result is
\beq
\fr{\de\nu}\nu=
\frac14(\ka_{HB})_{\rm lab}^{jj} 
-\frac14\hat N^j\hat N^k
[2(\ka_{DE})_{\rm lab}^{jk} 
+(\ka_{HB})_{\rm lab}^{jk}]\ .
\label{dnumicro2}
\eeq
Using transformation \rf{trans}
we express this in terms of
the Sun-frame coefficients.
We find
\beq
\fr{\de\nu}{\nu} =
-\frac14\hat N^j\hat 
 N^kR^{jJ}R^{kK}(\tilde\ka_{e'})^{JK}
-\half (\de^{jk}+\hat N^j\hat N^k)
R^{jJ}R^{kK}\ep^{JPQ}\be^Q (\tilde\ka_{o'})^{KP}
-\tilde\ka_{\rm tr}\ ,
\label{gendnu}
\eeq
where for convenience we
define the linear combinations
\beq
(\tilde\ka_{e'})^{JK}=
3(\tilde\ka_{e+})^{JK}
+(\tilde\ka_{e-})^{JK}\ ,
\quad
(\tilde\ka_{o'})^{JK}=
3(\tilde\ka_{o-})^{JK}
+(\tilde\ka_{o+})^{JK}\ .
\label{convenient}
\eeq
The matrix combinations
$\tilde\ka_{e'}$ and
$\tilde\ka_{o'}$ are traceless.
The first contains five
linearly independent
combinations of the 11
parity-even coefficients for
Lorentz violation,
while $\tilde\ka_{o'}$
contains all eight
parity-odd coefficients.

As an example, consider
two identical cavities,
operated in the above mode,
oriented at right angles to
each other on the ISS.
In general, the resonant
frequency of the cavities
will vary at the first and
second harmonics of the
stations orbital frequency $\om_s$.
A search for Lorentz violation
could be performed by looking
for this variation in the beat
frequency of the two cavities.
The variation takes the general form
\beq
\fr{\nu_{beat}}{\nu} \equiv 
\fr {\de\nu_1} {\nu} - \fr{\de\nu_2}{\nu}
= {\cal A}_s\sin\om_sT_s +{\cal A}_c\cos\om_sT_s 
+{\cal B}_s\sin2\om_sT_s+{\cal B}_c\cos2\om_sT_s + {\cal C} ,
\label{dnu7}
\eeq
where
${\cal A}_s$, ${\cal A}_c$,
${\cal B}_s$, and ${\cal B}_c$
are four linear combinations
of the coefficients
for Lorentz violation.
These combinations depend
on the orientations 
of the cavity pair and on the
orientation of the orbital plane 
with respect to the Sun-centered frame.
The precession of the ISS orbit
slowly changes the four combinations,
allowing access to more coefficients.
Typically, these are rather cumbersome
\cite{km}
and are omitted here.

The sensitivity to the coefficients
$(k_f)_{\ka\la\mu\nu}$ strongly
depends on the orientations of the cavities.
It can be shown that orienting
a cavity with $\hat N$
in the orbital plane
maximizes the sensitivity to
the second harmonics, at leading
order in $\be$.
Orienting a cavity so that
$\hat N$ is $45^\circ$ out of the
plane maximizes sensitivity to
the first harmonics.
A sensible configuration might 
have one cavity in the
orbital plane and one $45^\circ$
out of it.

There are many variations of the
above experiment that could be performed.
Earth-based experiments similar to
those discussed Sec. 4.2 could
also be performed using microwave cavities.
Operating in different modes or
using cavities filled with matter
changes the combinations of coefficients
to which the experiment is sensitive.
It is also possible to compare the
resonant frequency of a cavity
to a reference clock other than
another cavity oscillator. 
For example, the reference clock could be a
hydrogen maser or atomic clock,
which could conveniently be operated
on a transition known to be insensitive
to Lorentz violation
\cite{spaceexpt}.

With current stabilities,
it seems likely that microwave-cavity
oscillators could access coefficients
that are currently unmeasured,
at levels comparable to the those of
optical-cavity experiments and
perhaps at the $10^{-16}$ level.

\begin{table}[b!]
\begin{center}
\begin{tabular}{|l|c||c|c||c|c|}
\hline
\multicolumn{2}{|c||}{ }
& \multicolumn{2}{c||}{ Astrophysical Tests}
& \multicolumn{2}{c|}{ Cavity Tests}
\\
\hline
\hline
Coeff.\ & No.\ & Velocity & Polarization & Optical & Microwave \\
\hline
\quad
$\tilde\ka_{e+}$& 5 & -16 & -32  & $\star$ & - \\
\quad
$\tilde\ka_{e-}$& 5 & n/a & n/a & $\star$ & - \\
\quad
$\tilde\ka_{o+}$& 3 & n/a & n/a & $\star$ & - \\
\quad
$\tilde\ka_{o-}$& 5 & -16 & -32 & $\star$ & - \\
\quad
$\tilde\ka_{\rm tr}$& 1 & n/a & n/a & - & - \\
\hline
\end{tabular}
\end{center}
\caption{Existing constraints.}
\end{table}

\vspace{.4cm}\noindent
{\bf 5. Summary}\\[.2cm]
In this work, we considered
the experimental consequences
of a Lorentz-violating
electrodynamics which arises
from a Lorentz- and CPT-violating
standard-model extension.
We found that astrophysical
bounds on birefringence lead
to stringent constraints on
ten coefficients for Lorentz
violation.
Access to the remaining coefficients
may be accomplished through
clock-comparison tests involving
resonant cavities.

We summarize the current constraints
in Table 3.
The 19 coefficients
$(k_F)_{\ka\la\mu\nu}$
are represented by the 
matrices $\tilde\ka_{e+}$,
$\tilde\ka_{e-}$, $\tilde\ka_{o+}$,
$\tilde\ka_{o-}$, $\tilde\ka_{tr}$
defined in Eq.\ \rf{kappatilde}.
The number of independent
components in each matrix
is shown in the second column.
The third and fourth column
give the order of magnitude of
astrophysical bounds.
Laboratory experiments with
optical and microwave cavities
can in principle access
all the coefficients.
The matrices for which a
few components are probably
constrained by the optical
cavity experiments discussed in
Sec. 4.1 are  indicated by
the symbol $\star$ in the table.
To date,
no measurements of
Lorentz violation using
microwave cavities have
been reported. 

We conclude by remarking that
even though the ten coefficients
in $\tilde\ka_{e+}$ and
$\tilde\ka_{o-}$
are tightly constrained
by astrophysical measurements,
confirming these in laboratory experiments
provides an important check
because the systematics in the two
types of experiments are significantly
different.
Furthermore, cavity experiments
access currently unexplored
regions in parameter space, and they
offer the possibility of discovering
physics beyond the standard model.

\vspace{.4cm}\noindent
{\bf Acknowledgments}\\[.2cm]
I thank Alan Kosteleck\'y for
collaboration on this work.
The research described here
was supported in part by the 
National Aeronautics and Space Administration
under grant number NAG8-1770 and by the 
United States Department of Energy
under grant number DE-FG02-91ER40661.

\end{document}